\def\etal{{\it et al.\thinspace}}

\documentstyle[emulateapj,psfig,apjfonts]{article}

\begin{document}

\title{The formation and habitability of terrestrial planets in the presence of
hot jupiters}

\author{Sean N. Raymond\altaffilmark{1}, Thomas Quinn\altaffilmark{1}, \&
Jonathan I. Lunine\altaffilmark{2}}

\altaffiltext{1}{Department of Astronomy, University of Washington, Box 351580,
Seattle, WA 98195 (raymond@astro.washington.edu; trq@astro.washington.edu)}
\altaffiltext{2}{Lunar and Planetary Laboratory, The University of Arizona,
Tucson, AZ 85287. (jlunine@lpl.arizona.edu)}

\begin{abstract}

`Hot jupiters,' giant planets with orbits very close to their parent stars,
are thought to form farther away and migrate inward via interactions with a
massive gas disk.  If a giant planet forms and migrates quickly, 
the planetesimal population has time to re-generate in the lifetime of the
disk and terrestrial planets may form (Armitage 2003).
We present results of simulations of terrestrial planet formation in the
presence of hot jupiters, broadly defined as having orbital radii $\leq$ 0.5 AU.  
We show that terrestrial planets similar to those in the Solar System can form
around stars with hot jupiters, and can have water contents equal to or
higher than the Earth's.  
For small orbital radii of hot jupiters (e.g. 0.15, 0.25 AU) potentially
habitable planets can form, but for semi-major axes of 0.5 AU or greater their
formation is suppressed. 
We show that the presence of an outer giant planet such as Jupiter does not
enhance the water content of the terrestrial planets, but rather decreases their
formation and water delivery timescales.
We speculate that asteroid belts may exist interior to the terrestrial
planets in systems with hot jupiters.

\end{abstract}


\section{Introduction}

Roughly one third of the giant planets discovered outside the Solar System
have orbits within 0.5 astronomical units (AU) of their central
stars\footnote{Data from http://www.exoplanets.org}.
These ``hot jupiters''\footnote{We use the term hot jupiter to apply to gas
giant planets with semimajor axes $a \, \leq$ 0.5 AU.  This departs from
certain uses of the term, which reserve it for planets inside 0.1 AU.  In that
nomenclature, one might call planets at 0.5 AU ``warm jupiters.''}
are thought to have formed farther out and migrated inward via gravitational
torques with a massive gas disk (Lin, Bodenheimer \& Richardson 1996).  
If this migration occurs within the first million years (Myr) of the disk
lifetime, the planetesimal population (the building blocks of terrestrial
planets) is not strongly depleted.  However, if migration occurs later,
planetesimals are destroyed without enough time to re-form, making it
impossible for sizable terrestrial planets to form (Armitage 2003).

Recent results show that giant planets can form on very short timescales via
gravitational collapse (Boss 1997; Mayer \etal, 2002; Rice \etal, 2003).  
New simulations of the standard, core-accretion scenario (Pollack \etal, 1996)
including turbulence (Rice \& Armitage 2003) and migration during formation
(Alibert, Mordasini \& Benz 2004) have shown
that giant planets can form via this mechanism in 1 Myr or less, in agreement
with the observed, 1-10 million year lifetime of circumstellar disks
(Brice\~no et al 2001).
Observations of the $\sim$ 1 Myr old star Coku Tau 4 with the Spitzer Space
Telescope have revealed an absence of dust inside 10 AU.  One explanation is
the presence of a planet orbiting this very young star (Forrest \etal, 2004).
If correct, this would be observational evidence for fast giant planet formation.

The timescale for the inward migration of a giant planet depends on the mass
of the planet and the mass and viscosity of the gaseous disk, and
is typically less than 10$^5$ years for Saturn- to
Jupiter- mass planets (D'Angelo, Kley, \& Henning 2003).  Migration begins
immediately after, even during, the formation of the giant planet 
(Lufkin \etal, 2004).
The mechanism by which migration stops is not well understood, and may involve
interactions with magnetic fields (Terquem 2003) or an evacuated region in the inner
disk (Kuchner \& Lecar 2002; Matsuyama, Johnstone, \& Murray 2003).  Many
planets may in fact migrate all the way into the star (Nelson \& Papaloizou 2000). 

Based on the above arguments, we expect that terrestrial planets can form in a
standard, bottom-up fashion in the presence of a hot jupiter.

The character and composition of a system of terrestrial planets is strongly 
affected by the amount of solid material (Wetherill 1996; Chambers \& Cassen
2002; Raymond, Quinn \& Lunine 2004) and the presence of one or more giant
planets (Chambers \& Cassen 2002; Levison \& Agnor 2003; Raymond \etal, 2004).
The Earth acquired most of its water during formation from bodies which formed
in the outer asteroid belt, past the ``snow line,'' where water could exist as
ice in the low pressure protoplanetary disk (Morbidelli \etal, 2000). 

The habitable zone around a star is defined as the annulus in which the
temperature is right for liquid water to exist on the surface of an Earth-like
planet, and is roughly 0.95 - 1.37 AU in our Solar System (Kasting, Whitmire
\& Reynolds 1993). 
A potentially habitable planet not only needs to reside in its star's
habitable zone, it also needs a substantial water content.  The source of
water, however, lies much farther out in the protoplanetary disk, past the
snow line.   The formation of a habitable planet therefore requires
significant radial stirring of protoplanets with different compositions 
(see Raymond \etal, 2004 for a discussion). 

Here we present results of dynamical simulations of terrestrial planet
formation in the presence of a hot jupiter, both with and without an exterior
giant planet.  We include hot jupiters with orbital radii of 0.15, 0.25 and
0.5 AU, and in some cases outer giant planets at 5.2 AU.  Section 2 outlines
our initial conditions and numerical methods.  Section 3 presents our results,
which are discussed in section 4. 

\section{Method}

A simulation begins with a disk of protoplanets which reflects the
minimum mass solar nebula model (Hayashi 1981).  Planetary embryos have
densities of 3 $g \, cm^{-3}$ and are placed from the hot jupiter out to 5.2
AU.  These are randomly spaced by 3-6 mutual Hill radii assuming the surface
density of solids scales with heliocentric distance $r$ as $r^{-1.5}$.  
The surface density is normalized to 10 $g \, cm^{-2}$ at 1 AU,
with each disk of embryos containing 6-7 earth masses of material inside 5 AU.  
The discovered giant planets are found to preferentially orbit stars with
metallicities higher than the Sun's (Laws \etal, 2003), indicating that they 
likely contain a large amount of solid material 
with which to build terrestrial planets.  Our chosen value for the surface
density is therefore quite low, and accounts for some depletion
during hot jupiter migration.  All hot jupiters have masses of 0.5 Jupiter
masses and all outer giant planets are 1 Jupiter mass.

We assign protoplanets an initial distribution of water content which reflects
the distribution in chondritic meteorites (see Fig. 2 from Raymond \etal
2004), such that the inner bodies are dry, past 2 AU planetary 
embryos contain 0.1\% water, and past 2.5 AU embryos contain 5\% water.
Their iron distribution is interpolated between the content of the
planets and chondritic asteroid classes, ignoring the planet Mercury.  
These range from 0.40 (40\% iron by mass) at 0.2 AU to 0.15 at 5 AU.  
Each embryo is given a small initial inclination ($< 1^{\circ}$) and
eccentricity ($<$ 0.02).

Each simulation is evolved for at least 200 million years using a hybrid
integrator called Mercury (Chambers 1999), which evolves the orbits of all 
bodies and keeps track of collisions.
The hybrid scheme in Mercury uses a symplectic algorithm to evolve orbits of
bodies unless they are involved in a close encounter, in which case it
switches to a Bulirsch-Stoer method.  Collisions are treated as inelastic
mergers which conserve mass and water content. 
The time step in each simulation is chosen to be less than 1/20 of the orbital
period of the innermost body in the simulation, and ranges from 1 day for a
hot jupiter at 0.15 AU to 6 days for a hot jupiter at 0.5 AU.
Each simulation conserved energy to at least one part in 10$^5$, and took
between three weeks and two months to complete on a desktop PC.

\section{Results}

Figure 1 shows the time evolution of one simulation which formed a planet in
the habitable zone, with a hot jupiter at 0.25 AU and an outer giant planet at
5.2 AU (not shown).  Planetary embryos are dynamically excited by the giant
planets and their mutual gravitation, increasing their eccentricities and
causing their orbits to cross.  This results in both accretional impacts and
close encounters with giant planets, which eject roughly half of the
terrestrial bodies.  By the end of a simulation only a few terrestrial planets
remain. In this case four terrestrial bodies have formed including two planets
inside 2 AU, one of which lies in the habitable zone at 1.06 AU with 1.68
times the mass of Earth with water content higher than the Earth's. 
As our simulations do not account for water loss during impacts, water content
values are upper limits.  However, we do not simulate the secondary delivery
of volatiles from farther out in the disk (``late veneer'') which would
increase the water content.although likely not by more than 10\% if it
proceeds as in our Solar system (Morbidelli \etal, 2000).

Figure 2 shows the final state of twelve simulations (out of twenty), with the
Solar System included for scale.  Grey circles represent the positions of
giant planets in each simulation and are not on the same scale as the
terrestrial bodies.  The eccentricity of each body is shown beneath
it by its radial excursion over the course of one orbit.
Terrestrial planets can form in the habitable zone
in the presence of a hot jupiter, often with substantial water contents.
The possibility of a potentially habitable planet forming depends on the
location of the hot jupiter.  In most cases, no planets more massive than 0.2
Earth masses form within a factor of 3 in period to the hot jupiter, roughly a
factor of two in semi-major axis. If a planet forms in the habitable zone with
a hot jupiter at 0.5 AU, it is the innermost terrestrial planet and tends to
be relatively small and dry.  Water-rich planets form readily in the
habitable zone with a hot jupiter at 0.15 or 0.25 AU.

In our Solar System planet formation was suppressed in the asteroid belt by
the gravitational effects of Jupiter.  This is seen in Fig. 2, as no
terrestrial planets form within a factor of 3-4 in period to a hot jupiter or
an outer gas giant.  We speculate that as this gap is filled with the remnants
of terrestrial bodies in our solar system, systems with hot jupiters may
contain asteroid belts interior to the terrestrial planets.  The resolution of
current simulations is too low to test this hypothesis.

Simulations without a giant planet exterior to the terrestrial region form
planets of substantial mass in the asteroid belt and beyond on time scales
of hundreds of Myr.  Indeed, the systems in Fig. 2 with no outer giant planet
have not yet finished accreting.  Simulations 23 and 24 were run for 200 Myr,
sims 9 and 10 for 500 Myr, and sims 13 and 14 for 800+ Myr.  A comparison
between the outer regions of these demonstrates the long formation timescales.
An outer gas giant clears the asteroid belt of protoplanets quickly,
although the water content of terrestrial planets is roughly the same in the
absence of an outer giant planet.  In all cases, terrestrial planets in the
habitable zone form more quickly in the presence of an outer giant planet and
are delivered water at earlier times than with no outer gas giant.
This suggests that an outer giant planet's net effect
is to clear material from the asteroid belt and to accelerate terrestrial
planet formation.  Its role in delivering water
to the terrestrial planets is not a vital one in terms of quantity.

The amount of material ejected from the system is also a function of the
number and configuration of giant planets.  An outer giant planet ejects
approximately one half of the total
terrestrial mass in the system, while a hot jupiter can remove up to one
third.  In the case of a hot jupiter at 0.5 AU and an outer giant planet, the
terrestrial planets comprise only one quarter of the initial mass.  These
planets are systematically depleted in iron, because the inner, iron-rich
material has been largely removed by the hot jupiter.

We have run three simulations for one billion years or more to test the long
term stability of terrestrial planets in the presence of hot jupiters.  The
short dynamical timescales in the inner disk result in a fast clearing of
unstable objects, so a longer integration produces no change.  The asteroid
belt is slowly cleared by an outer giant planet, but all planets which are
well-separated from a giant planet (by a factor of 3-4 or more in orbital
period) are stable for long timescales.  

We have run two simulations under the assumption that the hot jupiter's
migration took place later in the lifetime of the protoplanetary disk.
The surface density of solid material was reduced by a
factor of five, and we included a hot jupiter at 0.25 AU and an outer giant
planet.  After 200 Myr of evolution, these systems formed no planets more
massive than 0.16 earth masses and left a large number of small bodies in the
terrestrial region, reminiscent of a large asteroid belt.

\section{Discussion}

All the simulations presented here contain giant planets on circular orbits
with fixed masses.  The observed hot jupiters inside 0.1 AU (51 Peg-type
hot jupiters) tend to have
circular orbits due to tidal interactions with the central star.  More distant
giant planets can have a large range in eccentricity and mass.  The effects of these
parameters can be extrapolated using previous results.  An eccentric giant
planet preferentially ejects water-rich material from the planetary system
rather than scattering it inward, which results in dry terrestrial planets
(Chambers \& Cassen 2002; Raymond \etal, 2004) with large
eccentricities, located far from the giant planet.  A more massive 
giant planet or a higher surface density of solid material results in a
smaller number of more massive terrestrial planets (Wetherill 1996; Chambers
\& Cassen 2002; Raymond \etal, 2004). 
We apply this to a known planetary system, 55 Cancri (Marcy \etal, 2002),
which contains two hot jupiters at 0.115 and 0.241 AU and an exterior giant
planet at 5.9 AU.
The hot jupiters are close to being in 3:1 resonance and the less massive one
has an eccentricity of 0.33.  The outer giant planet's eccentricity is 0.16
and it is four times as massive as Jupiter.  By our previous arguments, we
expect a small number of terrestrial planets to form in 55 Cancri far away
from the hot jupiters as well as from the outer giant.  The high
eccentricities should strongly deplete the solid material, resulting in
low-mass planets.  Simulations have shown this to be the case, with at most
two terrestrial planets forming in 55 Cancri, with masses no greater than 0.6
earth masses (Raymond \& Barnes 2004).

We have argued that terrestrial planets can form in the presence of hot
jupiters.  We have shown that potentially habitable planets with orbits in the
habitable zone and substantial water contents can form in such conditions.  We
hypothesize that asteroid belts may exist between the terrestrial planets and
a hot jupiter.
Based on this and previous work it is possible to predict the
character of the terrestrial planets around a star, from observables such as
the orbit and mass of a giant planet and the metallicity of the star.  Our
predictions will be testable in the near future with upcoming space missions
such as {\it Kepler}\footnote{http://www.kepler.arc.nasa.gov} and {\it
COROT}\footnote{http://www.astrsp-mrs.fr/projets/corot}, that will detect
giant and (hopefully) terrestrial planets around other stars.  Longer-term
missions like {\it Terrestrial Planet
Finder}\footnote{http://planetquest.jpl.nasa.gov/TPF} and {\it
Darwin}\footnote{http://ast.star.rl.ac.uk/darwin} hope to obtain spectra of 
terrestrial planets and search for signs of water and life.  
We suggest that stars with hot jupiters may be a good place to look for extra
solar terrestrial planets.

This result can also be applied to constrain the location of the Galactic
Habitable Zone (Gonzalez, Brownlee \& Ward 2002; Lineweaver, Fenner \& Gibson
2004).  This is defined as the region in the galaxy in which various factors
conspire to make the area suitable for life (e.g. the average metallicity of
stars, the rate of supernovae, time needed for life to evolve).  In
particular, Lineweaver \etal, (2004) assume (from Lineweaver 2001) that
the probability of a star to host a potentially habitable planet drops
precipitously if its metallicity is higher than 0.2-0.3 dex (solar metallicity
is defined to be 0.0).  This is based on the fact that higher metallicity stars
are more likely to have hot jupiters (Laws \etal, 2003), and the assumption
that any migration event would preclude the formation of terrestrial planets in
the system.  Our result, that potentially habitable planets {\it can}
exist around stars with hot jupiters, effectively widens the Galactic
Habitable Zone to include regions at small galactocentric distances and recent
times (``too metal rich'' regions in Figs. 3 and 4 of Lineweaver \etal, 2004).

\section{Acknowledgments}

We thank Lucio Mayer, Chris Laws, and Graeme Lufkin
for helpful discussions. This work was funded by NASA's Astrobiology Institute
and NASA Planetary Atmospheres.  These simulations were run under Condor,
which is publicly available at http://www.cs.wisc.edu/condor.


\begin{figure*}[h]
\centerline{\psfig{file=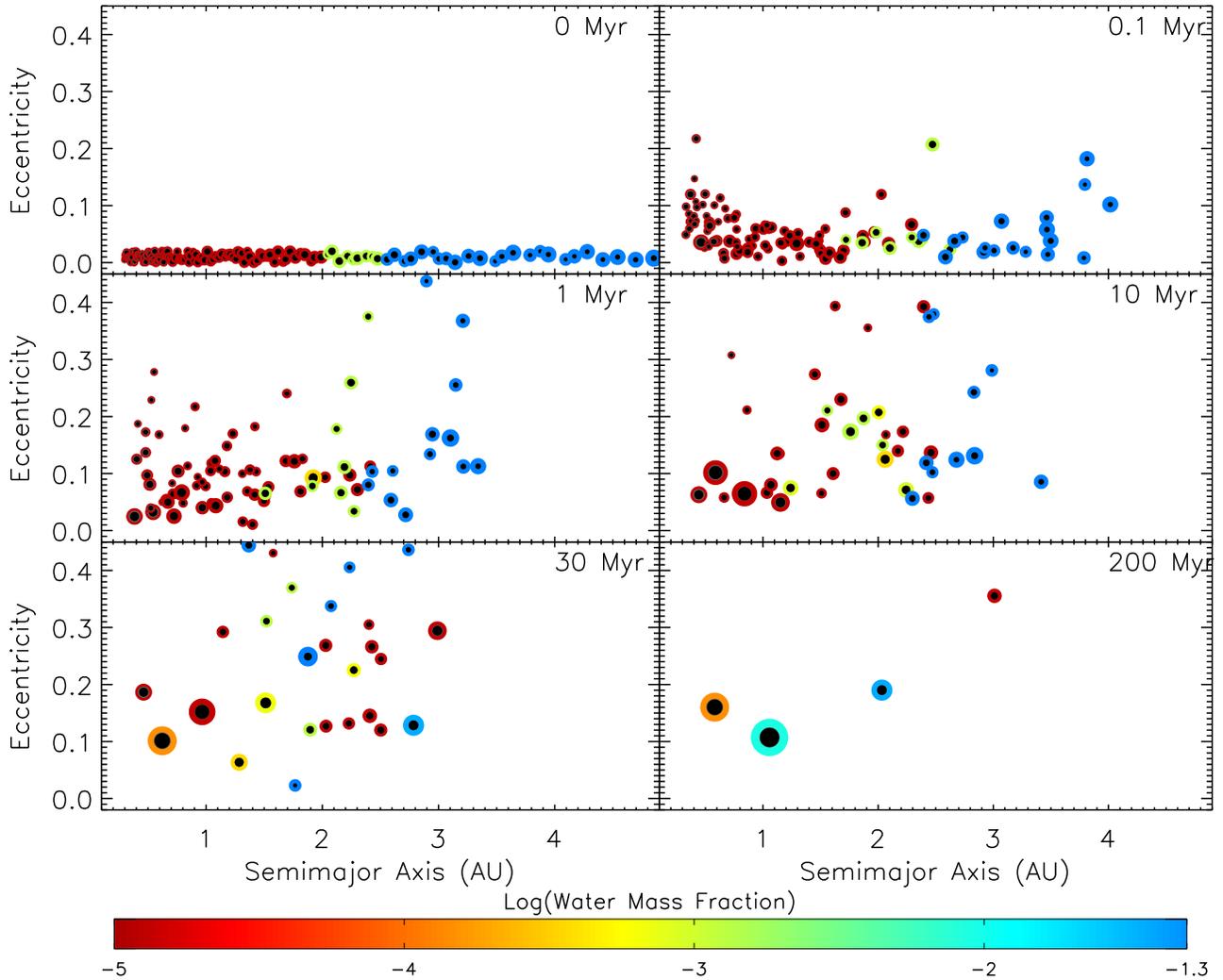}}
\caption{Six snapshots in time from a simulation with two giant planets
(not shown): a 0.5 Jupiter-mass hot jupiter at 0.25 AU and a Jupiter-mass
planet at 5.2 AU, both on circular, coplanar orbits.  Each panel plots the eccentricity
and semi-major axis of each surviving body in the simulation.  The size of a
body is proportional to its mass$^{(1/3)}$, and the dark region in the center
represents the size of its iron core, on the same scale.  The color
corresponds to the water mass fraction, which ranges initially from 10$^{-5}$
to 0.05.}
\label{fig:aet}
\end{figure*}

\begin{figure*}
\centerline{\psfig{file=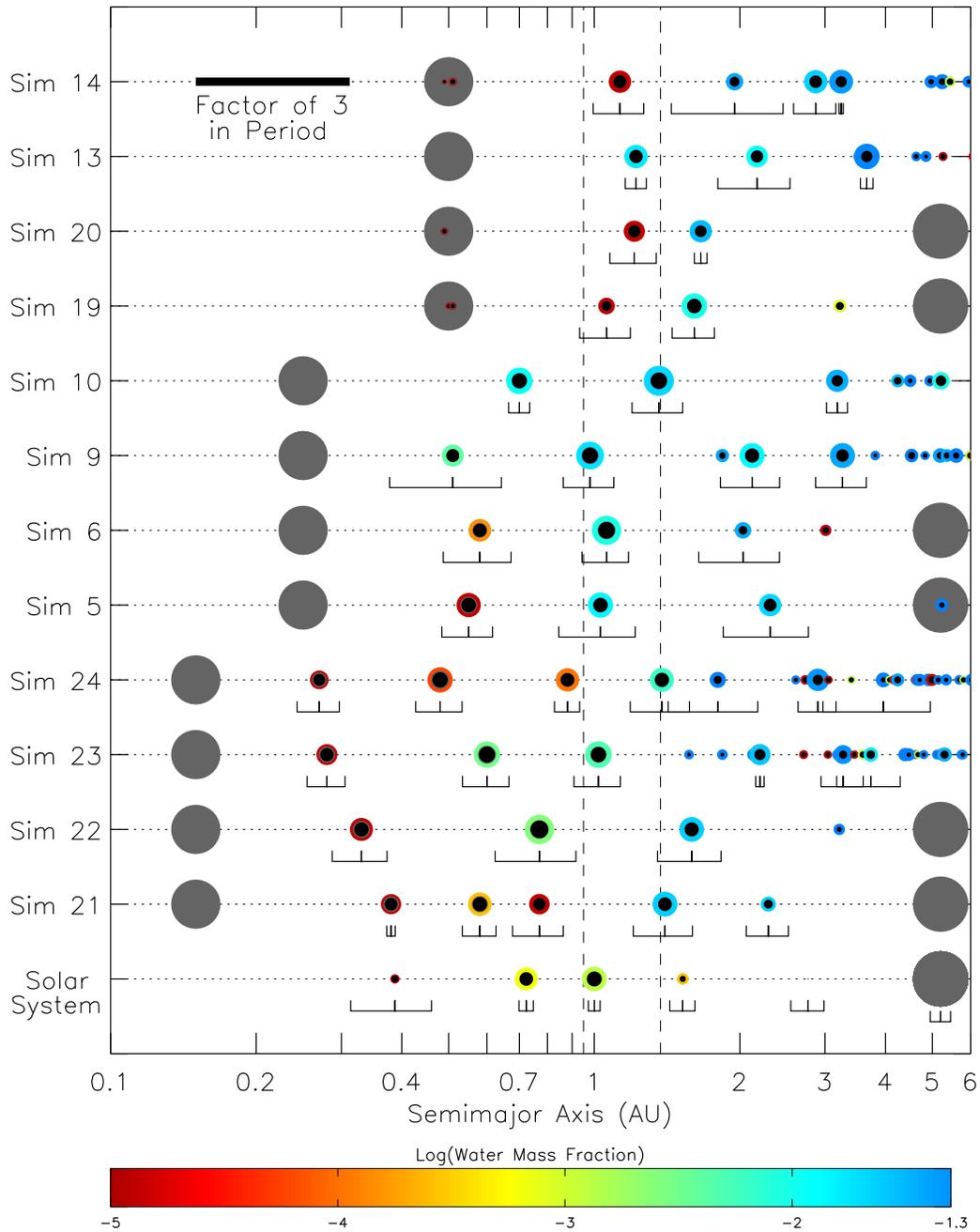,width=14cm}}
\caption{Final configurations of twelve simulations, with the Solar System
shown for scale.  The gray circles represent the giant planets in each
simulation and are not to the same scale as the terrestrial bodies.
The eccentricity of each body is represented by its excursion in heliocentric
distance over an orbit.  The x axis is on a logarithmic scale such
that a given separation corresponds to a fixed ratio of orbital periods, shown
in the scale bar on the top left.  The dashed vertical lines represent the
boundaries of the habitable zone (Kasting \etal, 1993).  Simulations 23 and 24 were 
run for 200 Myr, simulations 9 and 10 for 500 Myr, and simulations 13 and 14
for 800+ Myr.  A comparison shows the long accretion timescales in the
outer terrestrial region.  Note the presence of protoplanets in 1:1 resonance with
a giant planet in some cases.}
\label{fig:HJ7}
\end{figure*}

\end{document}